\documentclass[referee]{aa}  
\usepackage{graphicx}
\usepackage{txfonts}
\begin{document}
   \title{Numerical Simulation of Viscous-like Flow in and Around the Plasma Tail 
         of a Comet}

   \author{M. Reyes-Ruiz \inst{1}
          \and
          H. P\'erez-de-Tejada \inst{2}          
          \and
          H. Aceves \inst{1}
          \and
          R. V\'azquez \inst{1}
          }
          
   \offprints{M. Reyes-Ruiz}

   \institute{Instituto de Astronom\'{\i}a, 
  Universidad Nacional Aut\'onoma de M\'exico,
  Apdo Postal 877, Ensenada 22800, B.C., M\'exico \\
              \email{maurey, aceves, vazquez@astrosen.unam.mx, 
                     }
         \and
             Instituto de Geof\'{\i}sica, 
  Universidad Nacional Aut\'onoma de M\'exico,
  Ciudad Universitaria, C.P. 04510, M\'exico D.F. \\
             \email{perezdet@geofisica.unam.mx}
             }

   \date{Received September 15, 1996; accepted March 16, 1997}

 
  \abstract
   {}
   {To model the interaction of the solar wind with the plasma tail 
of a comet by means of numerical simulations, taking into account the 
effects of viscous-like forces.}
   {A 2D hydrodynamical, two species,  finite difference code has 
been developed for the solution  of the time dependent continuity, momentum and 
energy conservation equations, as applied to the problem at hand.}
   {We compute the evolution of the plasma of cometary origin in the 
tail as well as the properties of the shocked solar wind plasma around 
it, as it transfers momentum on its passage by the tail. Velocity, 
density and temperature profiles across the tail are obtained. Several 
models with different flow parameters are considered in order to study 
the relative importance of viscous-like effects and the coupling between species
on the flow dynamics. Assuming a Mach number equal to 2 for the incident 
solar wind  as it flows past the comet's nucleus, the flow exhibits three 
transitions with location and properties depending on 
the Reynolds number for each species and on the ratio of the timescale 
for inter-species coupling to the crossing time of the free flowing 
solar wind. By comparing our results with the measurements taken 
$in \ situ$ by the Giotto spacecraft during its flyby of comet Halley we 
constrain the flow parameters for both plasmas. }
   { In the context of our approximations, we find that our model is 
 qualitatively consistent with the $in \ situ$ measurements as long as 
the Reynolds number of the solar wind protons and of cometary H$_2$O+ ions is low, less 
than 100,  suggesting that viscous-like momentum transport processes 
 may play an important role in the 
interaction of the solar wind and the plasma environment of comets}

   \titlerunning{Viscous Flow in Comet Plasma Tails}
   \authorrunning{Reyes-Ruiz et al.}

   \maketitle
%

\section{Introduction}

The nature of the interaction of the solar wind with the plasma environment 
of comets as they approach the Sun, has been under investigation since the 
early days of space physics as a discipline (Biermann 1951, Alfv\'en 1957, 
see reviews by Cravens \& Gombosi 2004, and Ip 2004). The basic elements of the 
interaction were developed in the 20 years following the work 
of Biermann (1951), who proposed
that the interaction between the solar wind and the comet's plasma is 
responsible for the observed aberration angle of plasma tails with respect 
to the Sun-comet radius vector. Based on the inefficiency of Coulomb
collisional processes in the coupling of the solar wind and cometary plasmas,
Alfven (1957) proposed that the interplanetary magnetic field (IMF) is a 
fundamental ingredient in the solar wind-comet interaction; being responsible 
for channelling the cometary ions as it drapes into a magnetic tail. 
Biermann et al. (1967) suggested that as cometary ions 
are created and incorporated (picked-up) into the solar wind, the loading 
of the flow with this additional mass results in a modification of the flow 
properties as the solar wind approaches a comet; an idea further 
developed by Wallis (1973) (for a review see Szego 
et al. 2000). The IMF and mass loading are thus the main dynamical agents 
generally considered when 
developing models for the interaction of the solar wind with cometary 
ionospheres, as well as with other solar system bodies having an ionosphere
and without a strong intrinsic magnetic field. 

However, as has been pointed out by Perez-de-Tejada et al. (1980)  and
Perez-de-Tejada (1989), several features of the flow dynamics in the 
cometosheath and plasma tail of comets can be attributed to the action of 
viscous-like forces as the solar wind interacts with cometary plasma.
Such interaction processes are believed to be similar to those known 
to be occurring in other solar system bodies that have an ionosphere 
and no significant intrinsic magnetic field, particularly Venus and 
Mars (for a review see Perez-de-Tejada 1995, Perez-de-Tejada 
2009 and references therein). {\em In situ} measurements
indicate that, as in Venus and Mars, the solar wind flow in the ionosheath 
of comet Halley exhibits an intermediate transition, also called 
the ``mystery transition'', located approximately half-way between 
the bow shock and the cometopause (Johnstone et al. 1986, Goldstein et al. 
1986, Reme 1991, Perez-de-Tejada 1989 and references therein). 
Below this transition,
as we approach the cometopause, the antisunward velocity of the shocked solar 
wind decreases in a manner consistent with a viscous boundary layer 
(Perez-de-Tejada 1989). Also 
indicative of the presence of viscous-like processes is that the temperature of
the gas increases, and the density decreases, as we move from the intermediate
transition to the cometopause. Taking the distance between 
the intermediate transition and the cometopause as the thickness 
of a viscous boundary layer, which depends on the effective Reynolds number
of the flow ($R_{\rm eff}$), Perez-de-Tejada (1989) estimated that 
$R_{\rm eff} \approx 300$ for the solar wind flow in the cometosheath 
is necessary to reproduce the flow properties measured {\em in situ} by
the Giotto spacecraft on its flyby of comet Halley. 

An additional argument  suggesting the importance of 
viscous-like effects in the dynamics of the flow in the 
cometosheath and tail regions, follows from the comparison of 
the magnitude of the terms corresponding to momentum transport 
due to viscous-like forces and ${\bf J} \times {\bf B}$ forces in
the momentum conservation equation. Perez-de-Tejada (1999, 2000) 
has  argued
that downstream from the terminator in the ionosheath of Venus, a scenario 
analogous to the one considered in this paper, the fact that the flow is 
superalfvenic, as found from the {\em in situ} measurements of 
the Mariner 5 and Venera 10 spacecraft,  suggests that viscous-like forces 
may dominate over ${\bf J} \times {\bf B}$ forces in the flow dynamics
in the boundary layer formed in the interaction of solar wind and 
ionospheric plasma. If the flow is characterized by a low effective Reynolds 
number, $R_{\rm eff}$, this layer extends over a significant portion 
of the ionosheath of the planet. 

In comets, {\em in situ}
measurements obtained during the passage of the ICE spacecraft through the
tail of comet Giacobinni-Zinner (Bame et al. 1986, Slavin et al. 1986,
Meyer-Vernet 1986, Reme 1991) indicate that along the inbound trajectory 
(which lies slightly tailward of 
the comet nucleus) the magnetic field in the so-called transition and sheath 
regions, is approximately 10 nT, the number density is approximately 10 
cm$^{-3}$ and the tailward flow velocity varies from $\sim$ 
400 km/s (near the bow shock) down to 100 km/s. According 
to Perez-de-Tejada (1999), the ratio of viscous-like to magnetic forces is 
essentially the square of the alfvenic Mach number, 
$M_A^2 = V^2 / (B^2/8\pi\rho)$. 
From the data of the ICE spacecraft cited above, we find that $M_A^2$
ranges between 4 and 40 across the cometosheath and hence, viscous-like
stresses may dominate over ${\bf J} \times {\bf B}$ forces by a similar 
amount, or more, throughout the cometosheath region tailward of the nucleus.
In the vicinity of the plasma tail, the measurements of the ICE spacecraft 
(Bame et al. 1986, Slavin et al. 1986) indicate that the
midplane density, dominated by cometary ions, reaches values of 200 cm$^{-3}$ 
at the point where the magnetic field is a maximum 50 nT. With flow speeds 
of approximately 20 km/s, the square of the alfvenic Mach number reaches a 
minimum value of 2-3 so that, even in the plasma tail, 
viscous-like forces are, at least, as important as ${\bf J} \times {\bf B}$ forces following
the arguments of Perez-de-Tejada (1999). 

The fact that $M_A^2 >> 1$ in the cometosheath means that the magnetic 
energy density is much smaller than the kinetic energy associated with the 
inertia of the plasma. This implies that ${\bf J} \times {\bf B}$ forces
are not the dominant dynamical factor responsible for the large scale 
properties of the flow in the region. In fact, one can argue that 
the formation of a magnetic tail is an indication that 
in the cometosheath, the large-scale magnetic field does not dominate the dynamics, 
it is merely carried around by the superalfvenic flow. If the dynamics were controlled 
by the magnetic forces, field lines would not bend onto a magnetic tail and the 
direction of the ion tail would not be essentially in the direction of the local solar 
wind velocity. We believe that the magnetic field does play a crucial role in 
the momentum transfer between the solar wind and the cometary plasma, 
but it is the small scale, ``turbulent'' magnetic field component, that mediates 
the microscopic interaction between charged particles leading to the transfer of 
momentum that we are modelling as an effectively viscous process.

\subsection{On the origin of ``viscosity''}

The precise origin of the viscous-like momentum transfer processes invoked 
in the viscous flow interpretation of the intermediate transition, in 
the ionosheath of comet Halley and in other ionospheric obstacles to the 
solar wind, is not yet clear.  Typical properties of solar wind 
and cometosheath plasma result in a ``normal'' viscosity, as it appears 
in the Navier-Stokes equations when derived from Boltzmann's equation, 
that can be considered negligible in the flow dynamics. Using for example
properties of the shocked solar wind in the vicinity of the tail measured 
at comet Giacobini-Zinner, $n_i$ = 10 cm$^{-3}$, 
$|B| = 10$ nT and $T$ = 3 $\times$ 10$^5$  K (Bame et al. 1986, 
Slavin et al. 1986) one calculates the viscosity coefficient resulting 
from particle interactions according to Spitzer (1962, eqn. 5-55) to
be $\mu \sim 10^{17}$ g cm$^{-1}$ s$^{-1}$ . This extremely low value most 
likely represents a lower limit for the viscosity coefficient, since it 
reflects the ability to transport momentum across field lines in a plasma 
threaded by a strong, uniform magnetic field. A more appropriate expression 
for the plasma viscosity coefficient in the conditions of a cometosheath is 
probably given by the coefficient presented in Cravens et al. (1980), which 
corresponds to a plasma in a strongly fluctuating magnetic field. 
Perez-de-Tejada (2005) has calculated the viscosity coefficient for the 
ionosheath of Venus based on these results. If we use the same procedure 
to calculate the viscosity coefficient for the solar wind around the tail of 
a comet (with the conditions measured at Giacobini-Zinner) we 
find $\mu \sim 10^{-11}$ g cm$^{-1}$ s$^{-1}$. 

With typical values for  the solar wind velocity and mass density in the 
cometosheath around the tail of comet Giacobini-Zinner,  
$V$ = 200 km/s and $\rho = 1.67 \times 10^{-23}$ gm cm$^{-3}$ 
respectively (Bame et al. 1986), and adopting a characteristic 
lengthscale of 10$^5$ km for  the variation of the flow velocity 
(roughly the thickness of the sheath region), 
we find that the corresponding Reynolds number for the flow, based on 
the ``normal'' viscosity coefficient estimated above, is $Re > 10^5$.
This indicates that viscous effects resulting from the collisions between 
particles in this environment are negligible. Assuming that the Prandtl number 
is not very different from unity, as argued by Perez-de-Tejada (2005), we can 
also neglect heat conduction resulting from particle collisions.

However, as in Venus and Mars, strong turbulence 
has been measured in the ionosheath of comets Halley and Giacobinni-Zinner
(Baker et al. 1986, Scarf et al. 1986, Klimov et al. 1986, 
Tsurutani and Smith 1986) and, as it generally occurs in many fluid dynamics 
applications, turbulence is characterized (sometimes even defined) by a
dramatic increase in the efficiency of transport processes, viscosity 
included, in the flow (Lesieur 1990).  The likely importance of turbulent 
viscosity in this scenario is also expected in view of the large value of 
the Reynolds number estimated above. Also, as discussed by 
Shapiro et al. (1995) and  Dobe et al. (1999 and references therein) 
conditions in the ionosheath of Venus and Mars favour the development of
plasma instabilities leading to effective wave-particle interactions. 
If this mechanism operates also in the cometosheath, it may lead, as in 
these planets, to increased coupling between the solar wind and cometary plasma in 
a viscous-like manner as suggested by Perez-de-Tejada (1989). 
In our opinion this justifies a detailed
study of the hypothesis of viscous-like effects on the flow dynamics in 
solar wind-comet interactions. It is the purpose of this paper to begin 
these investigations. 

In this paper we present results of 2D hydrodynamical, numerical simulations 
of the flow of solar wind and cometary H$_2$O+ ions in the tail and tailward 
cometosheath of a comet. This is our first attempt to model the interaction 
of the solar wind with the plasma environment of a comet taking into account 
viscous-like forces which. We
review the estimation of the effective Reynolds number of
Perez-de-Tejada (1989), based on the comparison of {\em in situ} measurements
at comet Halley with results from 
numerical simulations of the viscous-like, compressible flow of the 
solar wind over a dense, cold and slow velocity gas representing the 
plasma tail of a comet. We also study the relative importance of 
viscous-like forces and the coupling between the fast moving protons of
the solar wind and the slow H$_2$O+ ions in the tail. We do the latter by comparing 
models with different values of the effective Reynolds number, 
the parameter controlling viscous-like effects, and the effective
coupling timescale between both species. 

The paper is organized as follows. In section 2 we present the formulation
of the problem, the basic equations, approximations and parameters. Section 3 presents 
results of a series of simulations with different model parameters. A 
comparison of our results with {\em in situ} measurements at comet Halley 
is discussed in section 4. Finally, in section 5 we 
summarize our main results and present our conclusions.

%

\section{Formulation of the problem}
\label{sec:problem}

We model the interaction of the solar wind with the plasma tail of a comet
using a 2D hydrodynamic, two species ($a$ and $b$), finite difference code 
that is an extension of the single species version presented in
Reyes-Ruiz et al. (2008).  Included in the dynamical equations is a coupling term 
between both species: solar wind protons and cometary ions, which we assume 
to be H$_2$O+ ions. This term allows the solar wind flow to get {\em mass loaded} 
with cometary ions as they diffuse upwards from the tail,  and cometary ions to
be accelerated by the fast, streaming solar wind. The coupling 
term is taken from the work of Szego et al (2000) who describe the treatment 
of mass loaded plasmas. However, in order to isolate the effects of 
the viscous-like forces, we do not consider the ongoing creation of new 
ions in the flow, by photoionization or any other mechanism, as is usually
done in mass loading studies. Considering that we are modelling only the 
flow in and around the tail of the comet, the only source of additional 
ions in our problem is through the boundary condition at 
the left hand edge of our simulation box (see \S 2.3). 
It is clear that the 2D character of our simulations is an approximation 
to the real problem and may not allow us to study
some processes that may be essential for the dynamical evolution of the flow.
We make this approximation considering that this is the first approach to 
the problem in which viscous-like forces are taken into account. We also neglect 
the effect of the IMF entrained in the solar wind flow, and leave for 
future work the study of the dynamical effects of ${\bf J} \times {\bf B}$ 
forces, although we do not expect these to be dominant in the region 
(see arguments in the Introduction section).

Since we are interested in the gas dynamics in the tail we focus on the 
region behind the coma, starting from a few times 10$^4$ km behind the 
comet's nucleus and extending downstream to a few times 10$^5$ km as 
illustrated in Figure 1. 


\begin{figure}
\resizebox{\hsize}{!}{\includegraphics{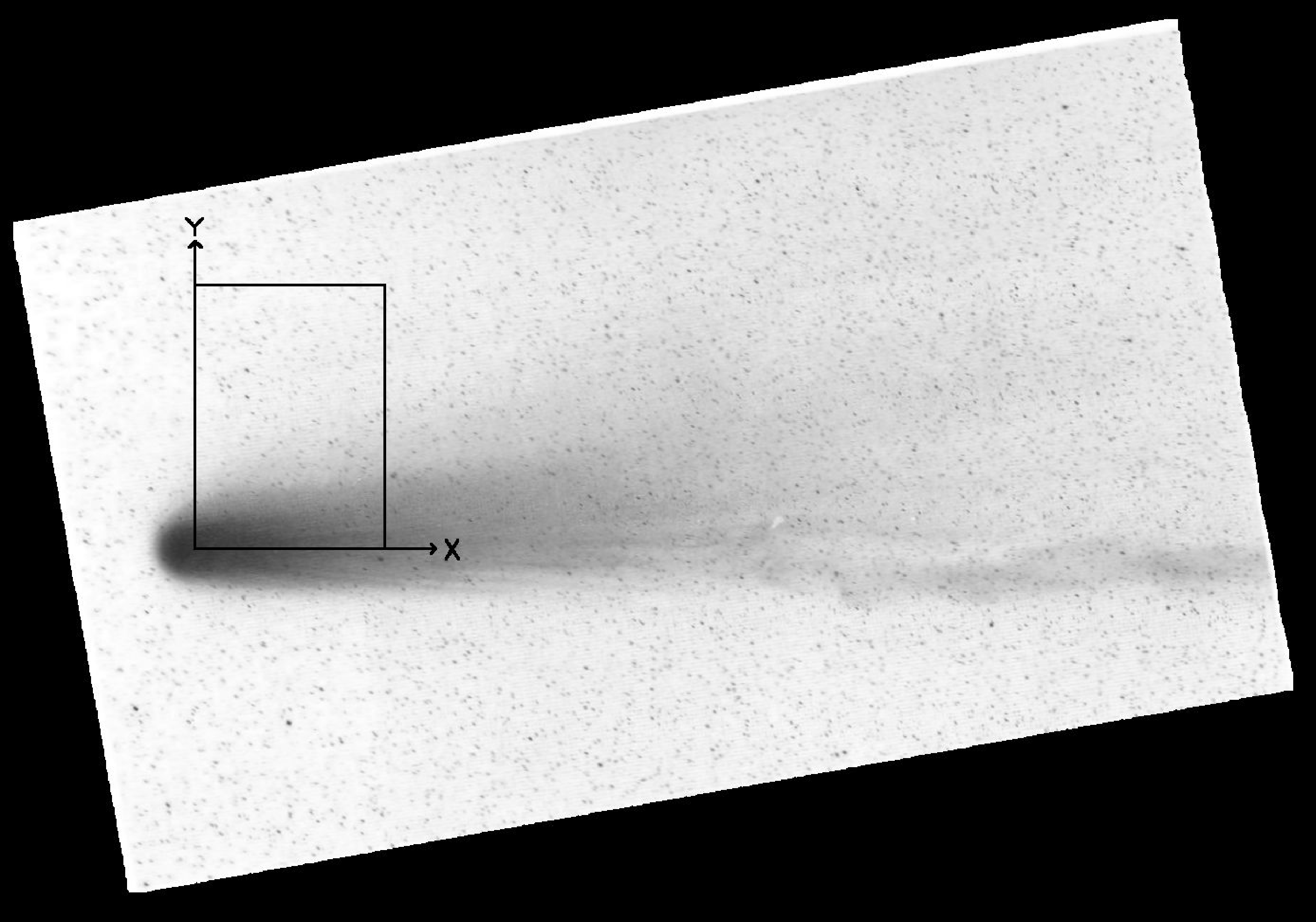}}
\caption {Illustration of the computational domain we use for 
our simulations. The box provides an approximate scale of the 
simulated region. Image of 
comet Halley taken the day of the encounter with Giotto by F.Miller, 
University of Michigan/CTIO (Brandt et al. 1992).}
\label{fig1}
\end{figure}


\subsection{Basic equations}

The present code solves the  Euler equations for mass, momentum 
and energy conservation, including terms representing the 
viscous-like effects and interspecies coupling due to turbulence and/or 
wave-particle interactions. In Cartesian 
coordinates and in conservative form, for species $a$, these can be written as:

\begin{equation}
\frac{\partial \vec{U}^a}{\partial t} + 
            \frac{\partial \vec{E}^a}{\partial x} + 
            \frac{\partial \vec{F}^a}{\partial y} = \vec{S}^{ab} 
      \label{eqnmot}
\end{equation}

\noindent where

\begin{equation}
\vec{U}^a = \left( \begin{array}{c}
  \rho^a \\
  \rho^a V_x^a \\
  \rho^a V_y^a \\
  E_t^a 
       \end{array} \right),             
\end{equation} 

\begin{equation}
\vec{E}^a = \left( \begin{array}{c}
  \rho^a V_x^a \\
  \rho^a V_x^a V_x^a + k_1^a p^a - k_2^a T_{xx}^a \\
  \rho^a V_x^a V_y^a - k_2^a T_{xy}^a \\
  (E_t^a + k_3^a p^a) V_x^a - 
     k_4^a (V_x^a T_{xx}^a + V_y^a T_{xy}^a) +
     k_5^a \dot{q}_x^a
       \end{array} \right) , 
       \label{Edef}           
\end{equation} 

\begin{equation}
\vec{F}^a = \left( \begin{array}{c}
  \rho^a V_y^a \\
  \rho^a V_x^a V_y^a - k_2^a T_{xy}^a \\
  \rho^a V_y^a V_y^a + k_1^a p^a - k_2^a T_{yy}^a \\
  (E_t^a + k_3^a p^a) V_y^a - 
     k_4^a (V_x^a T_{xy}^a + V_y^a T_{yy}^a) +
     k_5^a \dot{q}_y^a
       \end{array} \right),
         \label{Fdef}             
\end{equation} 

\noindent and the inter-species coupling term

\begin{equation}
\vec{S}^{ab} = \left( \begin{array}{c}
  O \\
  \rho^a \nu_{ab} (V_x^a - V_x^b )  \\
  \rho^a \nu_{ab} (V_y^a - V_y^b ) \\
  \frac{3}{2} k_3^a \rho^a \nu_{ab} \left[ T^b - T^a \right] + 
   \frac{k_3^a}{k_1^a} \rho^a \nu_{ab} \left[ {\bf V}^b - {\bf V}^a \right]^2
       \end{array} \right).  
         \label{Sdef}             
\end{equation} 

In the preceding equations $\rho^a$ is the mass density of gas $a$, $V_x^a$ and 
$V_y^a$ are its velocity components, $T^a$ is its temperature and $E_t^a$ is the total 
energy density of species $a$ defined by;

 It is important to point out that this form of the interspecies coupling 
term, although widely used in multispecies gas modelling in various astrophysical
scenarios (e.g. Schunk \& Nagy, 1980, Draine, 1986, Cravens, 1991, 
Falle, 2003, Van Loo et al. 2009, Szego et al. 2000 and references therein),
can be derived strictly from the Boltzmann collision integral 
only for the case corresponding to Maxwell molecules (see for example Gombosi, 1994). 
We use it for lack of a similarly simple, alternative expression for charged 
particles, and must be considered an approximation of uncertain validity in 
our case. Schunk (1977) has discussed the modifications to these expressions
for interspecies coupling for electrically charged molecules and in future
contributions we shall explore the effect of such modifications. In the 
present calculations we have chosen this approach to modelling multispecies 
flow, which follows the dynamics of each species separately, 
instead of an approach following a single fluid, composed of many 
different species, in order to clearly disentangle the widely different 
properties ($\rho, \ {\bf V}, \ T$, etc) of solar wind and cometary ions.
    
The coupling between species represented by the term $\vec{S}^{ab}$ in equation 
(\ref{eqnmot}) is taken from the work of Szego et al (2000),
and has the form of the traditional coupling resulting from binary 
collisions. The term $\nu_{ab}$ contained in $\vec{S}^{ab}$, reflects the
effective result of all processes able to transfer momentum and
energy from one species to another. Note that in the adimensional form 
of the equations $\nu_{ab}$ is actually $t_o \nu_{ab}$, that can be viewed 
as the ratio of the flow crossing time, $t_o = L/V_o$, to the inter-species
coupling timescale, $1/\nu_{ab}$. In order to preserve the symmetry between 
the coupling terms for both species, guaranteed by the identity 
$\rho^a \nu_{ab} = \rho^b \nu_{ba}$, we scale $\nu_{ab}$ as $\rho^b$ and 
$\nu_{ba}$ as $\rho^a$ with a single proportionality constant, $\nu_o$, 
which we take as uniform and constant. In our present code, $\nu_o$ 
enters as a parameter that can be varied to compare the importance of 
inter-species coupling to viscous-like forces.      
     
\begin{equation}
E_t^a = \rho^a e^a + \frac{1}{2} \rho^a (V^a)^2
\end{equation}

\noindent with $e^a$ being the internal energy per unit mass. In 
equations (\ref{Edef}) and (\ref{Fdef}), the coefficients $k_i^a$ ($i=1,5$) 
are the following 
combinations of dimensionless numbers and the adiabatic index for 
the gas, $\gamma^a$:

\begin{equation} 
k_1^a = \frac{1}{\gamma^a M_o^2},
\end{equation}
\begin{equation}
k_2^a = \frac{1}{R_{\rm eff}^a},
\end{equation}
\begin{equation}
k_3^a = (\gamma^a - 1),
\end{equation}
\begin{equation}
k_4^a = \frac{\gamma^a (\gamma^a-1) M_o^2}{R_{\rm eff}^a},
\end{equation}
\begin{equation}
k_5^a = \frac{\gamma^a}{R_{\rm eff}^a Pr^a},
\end{equation}

\noindent where the Mach number ($M_o$), the Reynolds number ($R_{\rm eff}^a$) and Prandtl 
number ($Pr^a$) for the flow of gas $a$, are defined respectively as:

\begin{equation} 
M_o = \frac{V_o}{C_{so}},
\end{equation}
\begin{equation}
R_{\rm eff}^a = \frac{\rho_o V_o L}{\mu_o^a},
\end{equation}
\begin{equation}
Pr^a = \frac{\mu_o^a c_p^a}{\kappa_o^a}.
\end{equation}

Quantities with subindex $o$ are those used for the 
normalization of the flow variables and parameters,  the reference 
sound speed is defined as $C_{so} = \sqrt{\gamma^a P_o /\rho_o}$,
$c_p^a$ is the 
specific heat at constant pressure for gas $a$, and $L$ is the normalization for 
the spatial coordinates. For simplicity we have assumed that the flow parameters, 
$\mu$ and $\kappa$, are uniform and that $\mu^a = \mu_o^a$, $\kappa^a = \kappa_o^a$,
$\mu^b = \mu_o^b$ and $\kappa^b = \kappa_o^b$.  
 
The terms $T_{xx}^a$, $T_{xy}^a$ and $T_{yy}^a$ in equations (\ref{Edef}) and 
(\ref{Fdef}) represent the components of the viscous-like stress tensor 
given by:

\begin{equation}
T_{xx}^a = \frac{4}{3} \frac{\partial V_x^a}{\partial x} -
         \frac{2}{3} \frac{\partial V_y^a}{\partial y} ,
\end{equation}

\begin{equation}
T_{xy}^a =  \frac{\partial V_y^a}{\partial x} +
          \frac{\partial V_x^a}{\partial y} , 
\end{equation}

\noindent and

\begin{equation}
T_{yy}^a = - \frac{2}{3} \frac{\partial V_x^a}{\partial x} +
         \frac{4}{3} \frac{\partial V_y^a}{\partial y} .
\end{equation}

 As is done in multiple fluid dynamics applications (Lesieur, 1990), we 
use the Boussinesq hypothesis in writing the Reynolds stress tensor, i.e.
we adopt a ``standard'' form for the relation between the viscous-like
stress tensor and the large scale flow velocity, using an effective viscosity 
coefficient that encapsulates turbulent viscosity as well as the possible 
effect of wave-particle interactions (Shapiro et al. 1995) 
or any other plasma 
instabilities leading to an increased coupling between ions in these 
collisionless plasmas. 

Also, in equations (\ref{Edef}) and (\ref{Fdef}), 
$\dot{q}_x^a$ and $\dot{q}_y^a$ 
are the components of the effective heat flux vector for species $a$
 (under the Boussinesq hypothesis):

\begin{equation}
\dot{q}_x^a =  - \frac{\partial T^a}{\partial x} ,
\end{equation}

\noindent and

\begin{equation}
\dot{q}_y^a =  - \frac{\partial T^a}{\partial y} .
\end{equation}
    
Furthermore, we have assumed throughout this work that both gases are 
ideal so that:

\begin{equation}
e^a = \frac{1}{\gamma^a -1} \frac{p^a}{\rho^a},
\end{equation}

\noindent with the equation of state, $p^a = \rho^a R T^a$. We have 
assumed  that both the solar wind plasma and the cometary plasma, 
in the tail region, 
are characterized by an adiabatic index,
$\gamma^a = \gamma^b = 5/3$. In the section 4 of the paper we present
some results for $\gamma^b = 1.25$, and discuss the effects of changing
this property of the cometary plasma.

An analogous set of equations and definitions are written for species $b$, and
both set of equations, coupled by the source term $\vec{S}^{ab}$ in equation 
(\ref{eqnmot}), are solved simultaneously.

\subsection{Numerical code}

The set of equations described above is discretized in space using 2nd order 
finite differences, 
and is advanced in time using an explicit, 2nd order MacCormack scheme
(Anderson 1995).  The implementation of the scheme is an extension of
that described in Reyes-Ruiz et al. (2008), but now with the additional source term
$\vec{S}$ in the equations of motion. In MacCormack's scheme the solution is 
advanced over one timestep
by a sequence of intermediate steps, the predictor
and corrector steps. In the predictor step an intermediate solution ($U^{\ast}$)
is calculated from the values of the physical variables, $\vec{U}_{i,j}^t$,
at a given time, $t$, and position, ($x_i$, $y_j$), according to:

\begin{equation}
\vec{U}_{i,j}^{\ast} =  \vec{U}_{i,j}^{t}   
     - c_1 \left[ \vec{E}_{i+1,j}^{t} - \vec{E}_{i,j}^{t} \right] 
     - c_2 \left[ \vec{F}_{i,j+1}^{t} - \vec{F}_{i,j}^{t} \right] 
     + \Delta t \ \vec{S}_{i,j}^t
\end{equation}


\noindent where $c_1 = \Delta t / \Delta x$, $c_2 = \Delta t / \Delta y$
and $\vec{E}^{t}$, $\vec{F}^{t}$  and $\vec{S}^t$are evaluated with $\vec{U}^{t}$
according to (\ref{Edef}), (\ref{Fdef}) and (\ref{Sdef}). This predicted 
solution is then corrected to obtain 
the solution at the next time, $t+\Delta t$, using:

\begin{eqnarray}
\vec{U}_{i,j}^{t + \Delta t} = \frac{1}{2} \left[ 
     \vec{U}_{i,j}^{t}  \right. &+&  \vec{U}_{i,j}^{\ast} 
     - c_1 \left( \vec{E}_{i,j}^{\ast} - \vec{E}_{i-1,j}^{\ast} \right) 
      \nonumber \\
     & - &  c_2 \left( \vec{F}_{i,j}^{\ast} - 
     \vec{F}_{i,j-1}^{\ast} \right) 
     + \left. \Delta t \ \vec{S}_{i,j}^{\ast}
     \right]
\end{eqnarray}
 
\noindent where $\vec{E}^{\ast}$, $\vec{F}^{\ast}$ and $\vec{S}^{\ast}$ 
are computed from $\vec{U}^{\ast}$ using (\ref{Edef}), (\ref{Fdef})
and (\ref{Sdef}). Further details of the implementation of MacCormack's
scheme are given in Reyes-Ruiz et al. (2008). 

A final upgrade to our previous code 
is the ability to handle some types of non-uniform, cartesian grids. For 
the simulations done in this work, the 
grid is defined by a series of ($x_i$, $y_j$) coordinates for which the
spacing is arbitrary. In our simulations the $x_i$ points are geometrically
distributed from $x_{min}$ to $x_{max}$ with $nx$ elements. The $y_j$
points are equispaced at the initial location of the tail (from $y = 0$ 
to $y = 1$ having 30 gridpoints) and geometrically distributed 
from $y = 1$ to $y = y_{max}$.
In both series the common ratio is 1.02. The 2nd order approximation for the 
$x$-derivative of a function $f$ at $x_i$ can be easily obtained from the 
Taylor series expansion of the function at $x_{i-1}$  and $x_{i+1}$, and 
is given by: 

$$
\left(\frac{df}{dx}\right)_i = 
\frac{\Delta x_{i-1}^2 f_{i+1} + 
       [\Delta x_{i}^2 - \Delta x_{i-1}^2] f_i - \Delta x_{i}^2 f_{i-1}}
      {[\Delta x_{i-1} \Delta x_{i}^2 - \Delta x_{i} \Delta x_{i-1}^2]}
$$

\noindent where $\Delta x_{i} = x_{i+1} - x_{i}$. An analogous expression 
exists for the $y$-derivative. This grid allows a higher 
resolution in the vicinity of the region of strong interaction,
while putting the $y=y_{max}$ boundary sufficiently far to avoid numerical
artifacts in our results.

\subsection{Initial and boundary conditions}

The solution for the flow is evolved from the following initial conditions. 
A dense, cold, slow moving plasma representing the tail is located between 
$y=0$ and $y=1$. Both H$_2$O+ and H+ ions are present in the tail, but with
protons much less abundant than H$_2$O+. Between $y=1.5$ and $y=y_{\rm max}$, 
the gas has the properties of a shocked, hot, fast moving solar wind
that contains both H$_2$O+ and H+ ions, with the number density of protons  
50 times greater than H$_2$O+. In all the calculations presented 
here, we have adopted a value $M_o = 2$ for the Mach number of the shocked 
solar wind incident on our computational domain. This assumption is made
based on the results of Spreiter \& Stahara (1980) who computed the 
the gas dynamics of the flow of the shocked solar wind in the ionosheath of 
Venus. Spreiter \& Stahara (1980) found that the flow is characterized 
by $M = 2$, as the solar wind crosses the terminator of the planet (the 
line separating the day and night sides) and 
heads tailwards.  In comets, we take the terminator to coincide approximately 
with the location of the nucleus.       
Between $y=1.0$ and $y=1.5$ there is 
transition region where the flow properties change smoothly in an exponential 
manner from those in the tail to those in the solar wind. The initial density 
of each species is taken to be:

\begin{equation}
\rho^{a}(t=0) = \left\{ \begin{array}{rl}
  0.025 \rho_{\rm tail} &\mbox{ if $y<1$} \\
  \rho_{\rm sw} &\mbox{ if $y>1.5$}
       \end{array} \right. \label{BCda} ,
\end{equation}
\begin{equation}
\rho^{b}(t=0) = \left\{ \begin{array}{rl}
  \rho_{\rm tail} &\mbox{ if $y<1$} \\
  0.32 \rho_{\rm sw} &\mbox{ if $y>1.5$}
       \end{array} \right. \label{BCdb} .
\end{equation}

We assume both species are moving initially with the same velocity: 

\begin{equation}
V_x^{a,b}(t=0) = \left\{ \begin{array}{rl}
  V_{\rm tail} &\mbox{ if $y<1$} \\
  V_{\rm sw} &\mbox{ if $y>1.5$}
       \end{array} \right. \label{BCvx}  , 
\end{equation}

\begin{equation}
V_y^{a,b}(t=0) = 0.\label{BCvy}
\end{equation}

In normalized quantities $V_{\rm sw} = 1$ and $V_{\rm tail}= 0.01$. 
For the results shown here we use, in normalized variables, 
$\rho_{\rm sw} = 1$ and $\rho_{\rm tail}= 400$. The local temperature 
of both species is assumed the same inside the tail, with  
$T_{\rm tail}^a = T_{\rm tail}^b = T_{\rm tail}$ and 
$T_{\rm sw} = 100 T_{\rm tail}$, with $T_{\rm sw} = 1$ in normalized
units. Outside the tail, for $y > 1.5$, cometary ions are injected with 
a temperature an order of magnitude lower than the streaming solar
wind protons, $T_{\rm sw}^a = 10 T_{\rm sw}^b = T_{\rm sw}$. 
This choice of temperatures and densities is made to yield an initial 
pressure balance between the H$_2$O+ plasma (species $b$) inside the
tail, and the proton plasma (solar wind, species $a$) outside.
With $p^a = \rho^a T^a$ and $p^b = ( m_p^a / m_p^b) \rho^b T^b$, 
$m_p^a$ and $m_p^b$ being the particle mass for species $a$ and $b$
respectively, we find that our choice of initial conditions
is characterized by a pressure in-balance among each species.   
Whether the rapid movement of cometary ions resulting from 
this initial condition is prevented by the wrapped-around IMF over
the comet's tail will be the subject of future studies.  
Although significantly different from the flow properties
at later times in the simulations, for all the cases we have
studied these 
initial conditions do not give rise to any long lasting instability 
in the flow so that the final state does not depend on their
precise form or value. 


\begin{figure}
\resizebox{\hsize}{!}{\includegraphics{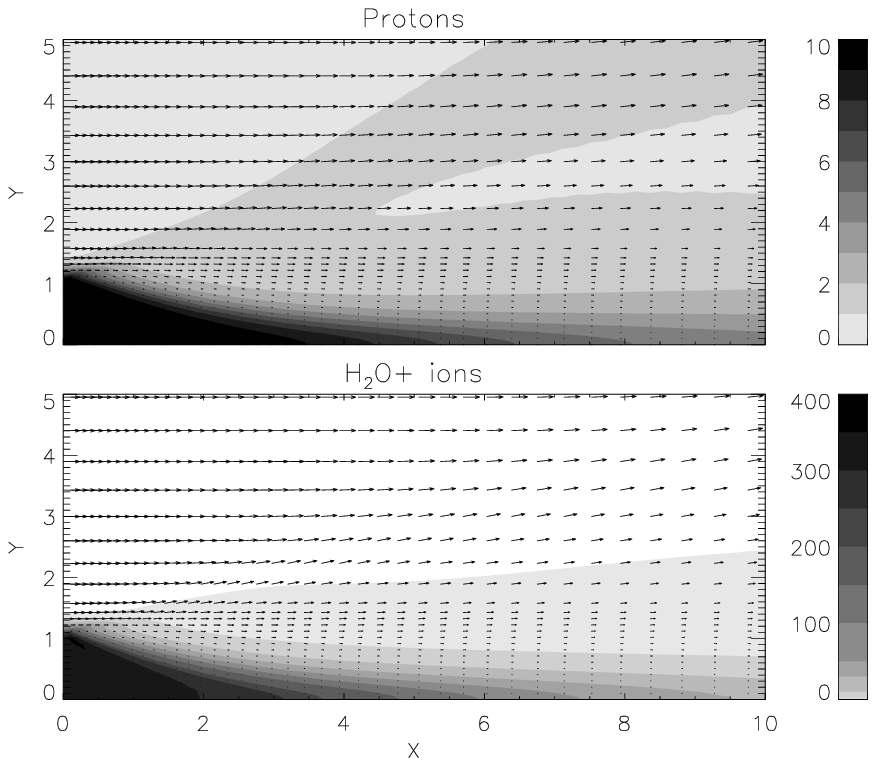}}
\caption {Density contours (shades of gray) and flow geometry (velocity vectors) 
for Case 1 ($Re^{a,b} = 30$, $\nu_o = 0.1$) after 1234 simulation time units. 
The top panel shows the configuration for the proton plasma (species $a$) and 
the right side panel shows the ``equilibrium'' configuration for cometary 
H$_2$O+ ions. Density and velocity are in normalized units.}
\label{fig2}
\end{figure}


The boundary conditions are chosen to be consistent with the 
initial conditions. At the left boundary, $x=x_{min}$, the flow 
density and velocity follow exactly that given by the initial condition 
in equations (\ref{BCda})-(\ref{BCvy}). Considering that the 
inflow to the comet's tail ($y<1$) is subsonic, we allow the 
inflow pressure to float freely as a linear extrapolation of the 
active mesh values (e.g. Anderson 1995). The right side boundary, 
$x=x_{max}$, corresponds to the commonly used outflow conditions for 
supersonic flows, namely the derivatives being zero for all flow variables.
We have also run simulations with an outer boundary condition 
obtained from linearly extrapolating the flow variables, resulting 
only in minor differences 
in the last gridpoints before the $x=x_{max}$ boundary. 


\section{Results}

We have performed a series of simulations with different set of parameters 
$R_{\rm eff}^{a,b}$ and $\nu_o$ to determine the effect of viscous forces and
inter-species coupling in the flow dynamics. 
For all cases considered, the flow evolves 
from the prescribed initial condition [eqns. (\ref{BCda})-(\ref{BCvy})], 
passing through a fast transient phase, during which a considerable portion
of the mass originally in the tail is eroded by the solar wind exiting 
our simulation domain. The relevance of this transient phase, lasting a few 
tens of solar wind crossing times ($t_o = L/V_o$), in relation to 
observed features in the evolution of the ion tail, will be
analysed in a future publication. In this work we concentrate on the 
following, quiescent stage of evolution since, given its longer timescale 
for existence, is more likely to be encountered.
In all cases, we present results for the flow velocity, density and 
temperature after a time long enough that a quasi-steady state 
has been reached.   All results are presented in terms of normalized 
quantities as defined in the previous section. For a particular application, 
appropriate values of $L$, $V_o$, $\rho_o$ and $T_o$ can be chosen as 
exemplified in Section 4 for comet Halley.

%
%
\begin{figure*}
\centering
\includegraphics{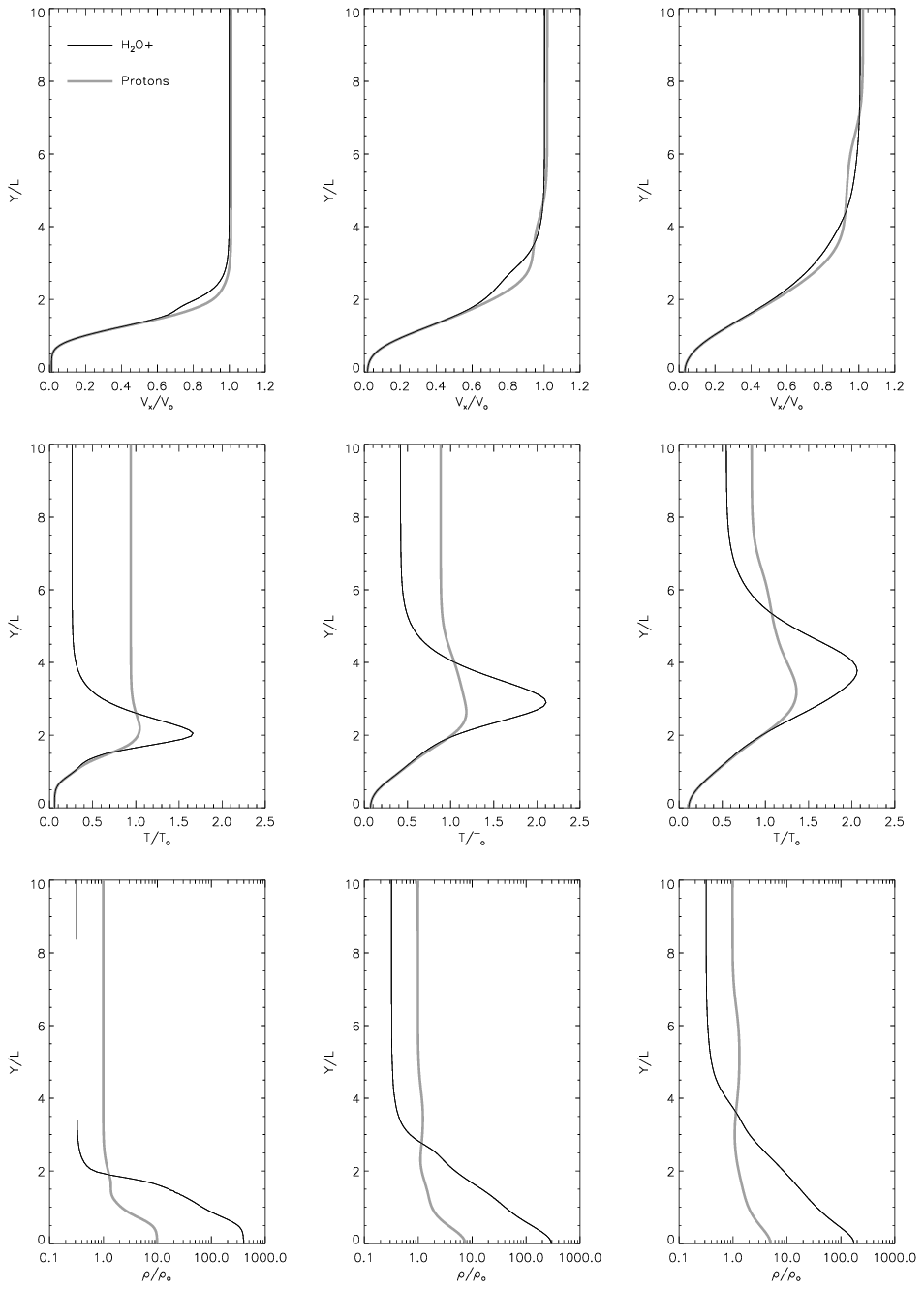}
\caption {Vertical profiles of the flow properties for Case 1 at three different 
positions; $x = 2$ (left column of panels), $x=5$ (middle column) and 
$x=8$ (right column). In all cases, gray lines indicate the properties 
of the proton plasma (species $a$) and black lines denote the properties of 
the H$_2$O+ plasma (species $b$). The top row shows the $x$ component of 
velocity, $V_x^{a,b}$,
the middle row shows the temperature, $T^{a,b}$, and the bottom row shows the 
mass density, $\rho^{a,b}$. All quantities are in normalized units.}
\label{fig3}
\end{figure*}
%
%

To determine the appropriate value of the effective Reynolds number for each species 
we consider the following. According to Perez-de-Tejada (1989) 
the geometry of the flow, measured 
{\em in situ} by the Giotto spacecraft 
in its fly-by comet Halley in march 1986, implies an effective
Reynolds number around 350 for the shocked solar wind flow above the 
cometopause along the spacecraft trajectory. In contrast, in a 
similar region in the ionosheath of Venus, Perez-de-Tejada (1999) 
and Reyes-Ruiz et al (2008) estimate a value of the Reynolds 
number an order of magnitude smaller ($R_{\rm eff}$ = 20), based on  
a comparison of {\em in situ} measurements (by the Venera 10
and Mariner 5 spacecraft) at Venus with the flow properties 
derived from a numerical simulation of the viscous-like solar wind-ionosphere
interaction. To assess the estimation of Perez-de-Tejada (1989) we 
have conducted simulations with 3 different values of the Reynolds 
number. A high value, $R_{\rm eff}$ = 100, similar to that estimated 
by Perez-de-Tejada (1989) for comet Halley; an intermediate value,
$R_{\rm eff}$ = 30, comparable to the value estimated by Reyes-Ruiz et al. (2008)
for the solar wind flow in the ionosheath of Venus; and a low value,
$R_{\rm eff}$ = 10, used to verify the tendency in the results 
as $R_{\rm eff}$ is decreased.    

In most cases, the value of the effective Reynolds number for both species is 
assumed to be the same, In our view, the lack of knowledge of the precise mechanisms
giving rise to the effective viscosity in these plasmas justifies
this assumption. However, we have analysed a case with different 
values of the effective Reynolds number for each species in the Discussion section. 

The value of $\nu_o$ is also varied to explore the relative importance of 
inter-species coupling, $versus$ viscous forces, which are
proportional to $R_{\rm eff}^{a,b}$. We will show results for 3 different cases:
a strong coupling case characterized by $\nu_o$ = 1, which can be interpreted as
having the timescale for inter-species coupling equal to the solar wind crossing 
timescale, $t_o = L/V_o$; a medium coupling case, in which the coupling timescale 
is an order of magnitude greater than the crossing timescale, $\nu_o$ = 0.1; and 
a weak coupling case, for which $\nu_o$ = 0.01, so that inter-species coupling 
effects are much smaller than other dynamical effects. 

To compare the state 
of the flow at the same time in its evolution for all cases, 
starting from the same initial condition, we have chosen,
arbitrarily, to show results at $t$ = 1234, 
with time units in multiples of the solar wind crossing time. The number of 
timesteps required to reach this time depends on the model parameters, 
for most cases less than 200 000 timesteps are required.   

%
%
\begin{figure}
\centering
\includegraphics{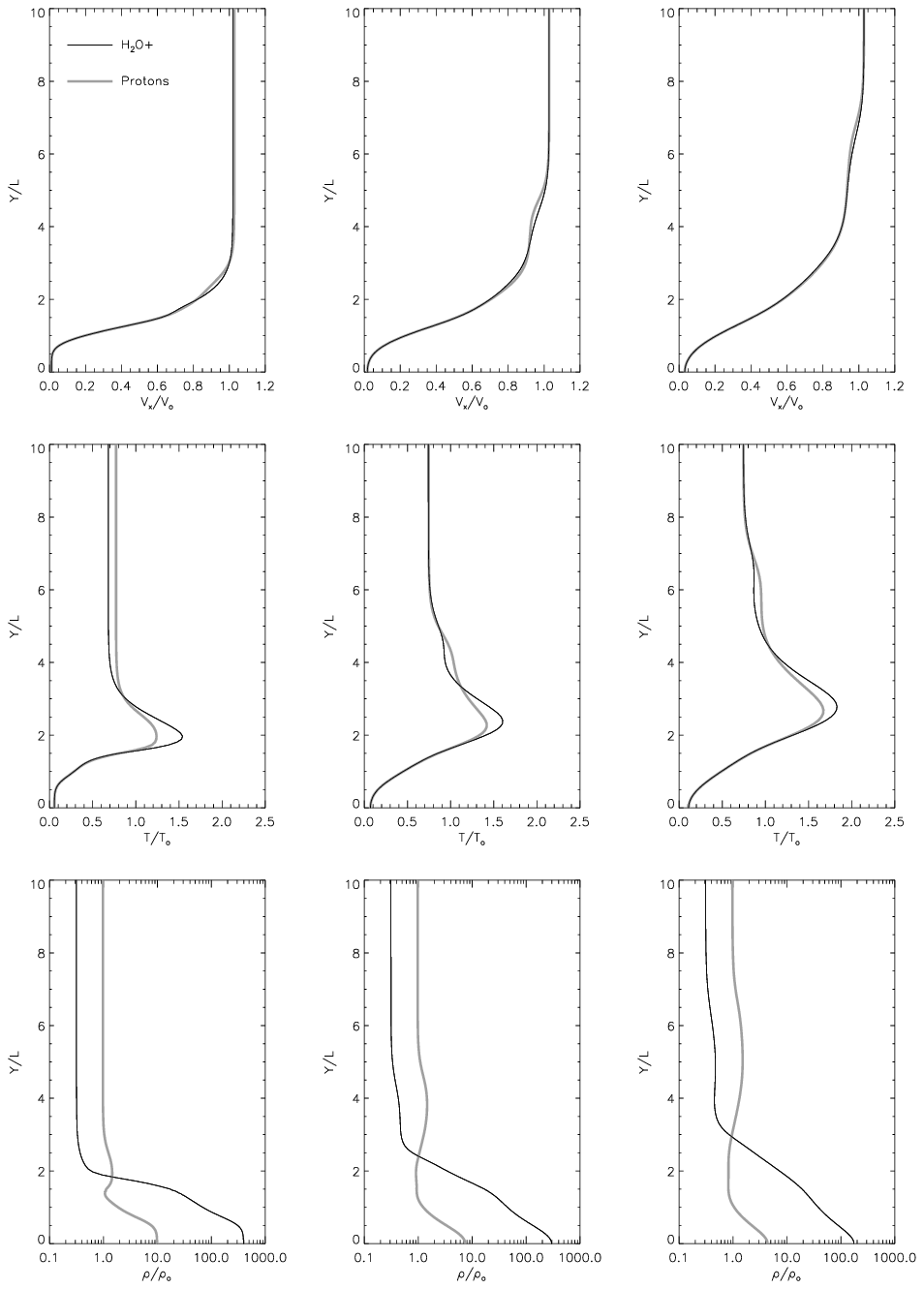}
\caption {Same as in Figure 3 but for simulation Case 2 
($R_{\rm eff}^{a,b} = 30$, $\nu_o = 1.0$).}
\label{fig4}
\end{figure}
%
%

\subsection{Effect of inter-species coupling}

Our fiducial model, Case 1, is characterized by model parameters 
$R_{\rm eff}^{a,b}$ = 30 and $\nu_o$ = 0.1.
In Figure \ref{fig2} we show density contours and the flow velocity for each species.
Initially the tail contained a uniform density, $\rho^a$ = 10 (protons) 
and $\rho^b$ = 400 (H$_2$O+ ions) for $y < 1$, and after $t = 1234$ $t_o$, 
a significant portion of 
the tail has been eroded by the effect of viscous forces and inter-species coupling. 
A shock wave is evident in the deflection of the flow velocity from the initial 
uniform distribution imposed by the boundary condition at $x = x_{\rm min}$. Also 
noticeable is the strong velocity gradient around $y = 2$ which corresponds to  
the viscous boundary layer. Both effects are
also shown in Figure \ref{fig3}, where vertical profiles of the $x$ component of
velocity, $V_x$, temperature, $T$, and mass density, $\rho$, are shown for three 
different $x$-positions, $x = 2, \ 5$ and $8$; in the left, middle and right
columns of each figure, respectively.    

In Figure \ref{fig3} the shock front and the boundary layer can be identified 
at all 3 
positions, but they are well separated only for $x = 5$ and $x = 8$, shown
in the middle and right hand columns, respectively. For a given $V_x$ profile, 
the shock front corresponds to the uppermost 
decrease from the uniform velocity ($V_x = 1$) in the free flowing solar wind. In the 
middle panel, corresponding to $x = 5$, this transition is located approximately 
at $y = 5$. A second transition, located approximately at $y = 2.5$ for $x = 5$, 
marks the top of the viscous boundary layer, below which the velocity drops sharply 
to the very low flow velocities in the middle of the tail. The shock front can 
also be seen as an increase in both temperature and density in the corresponding 
panels for each position. The temperature increase and density 
decrease characteristic of viscous boundary layers and found in previous studies 
of viscous flow over a flat plate (e.g. Reyes-Ruiz et al. 2008),
is also observed in other cases modeled here. This clearly indicates
that the region around $y = 2$ (at $x = 5$) is indeed a viscous boundary
layer. 

%
%
\begin{figure}
\centering
\includegraphics{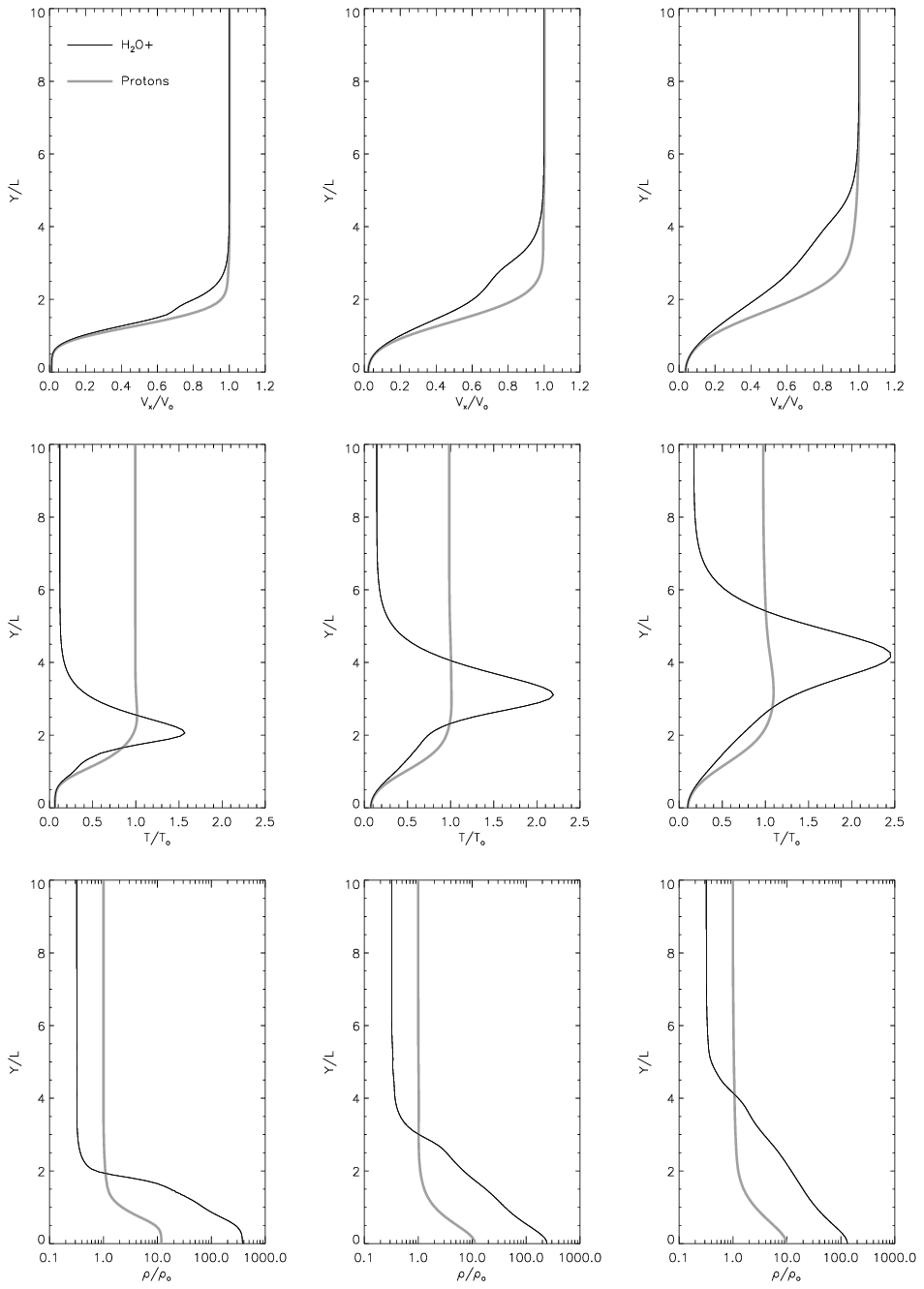}
\caption {Same as in Figure 3 but for simulation Case 3
($R_{\rm eff}^{a,b} = 30$, $\nu_o = 0.01$).}
\label{fig5}
\end{figure}
%
%

For Case 2 we use the same Reynolds number, $R_{\rm eff}^{a,b}$ = 30, as in Case 1, but 
increase the importance of inter-species coupling by using $\nu_o$ = 1.0. A Figure 
showing the general flow geometry is not shown since no 
appreciable differences are found with Case 1 (shown in Figure \ref{fig2}). 
However, the vertical profiles of flow 
properties, shown in Figure \ref{fig4}, clearly illustrate 
the effect of a much stronger inter-species coupling used in this model. Namely,
as both species are more tightly coupled, their velocity and temperature 
distribution tend to be almost identical. The density distribution 
conforms to the different boundary conditions for each species, since these 
are different, there is no reason why both densities should tend to 
equalize and they do not. Figure \ref{fig4} shows that the 
shock front and the boundary layer 
are not well separated at the rightmost position shown, $x = 2$. From the shock 
front height and boundary layer thickness shown in the middle and right columns 
of Figure \ref{fig4}, we see that both are proportional to the 
inter-species coupling (see \S4). The shock front height at $x = 5$, for example, 
changes from $y = 4.7$ for Case 1, to $y = 5.5$ 
in this case, while the thickness of the boundary layer goes from $y = 2.8$ to $y = 3.2$
as we increase the inter-species coupling parameter from 0.1 to 1. 

In Figure \ref{fig5} we show the results for Case 3 characterized by a very weak 
inter-species coupling, $\nu_o = 0.01$. The general 
flow geometry (not shown) is very similar to that in Figure \ref{fig2}. A comparison  
of Figure \ref{fig5} (weak coupling) with Figures \ref{fig3} and \ref{fig4} 
(medium and strong coupling respectively), clearly shows that in Case 3 the dynamics of 
both species is essentially uncoupled. The location of the shock front and the 
top of the boundary layer are different for each species. For example, 
at $x = 5$, only for the cometary H$_2$O+ ions the shock front and boundary layer
are clearly separated. For the H$_2$O+ ions the shock front is located 
approximately at $y = 5$ and the top of the boundary layer is at $y = 2.5$, 
while for protons the shock front and the top of the boundary layer are
both located around $y = 2$.      

%
%
\begin{figure}
\resizebox{\hsize}{!}{\includegraphics{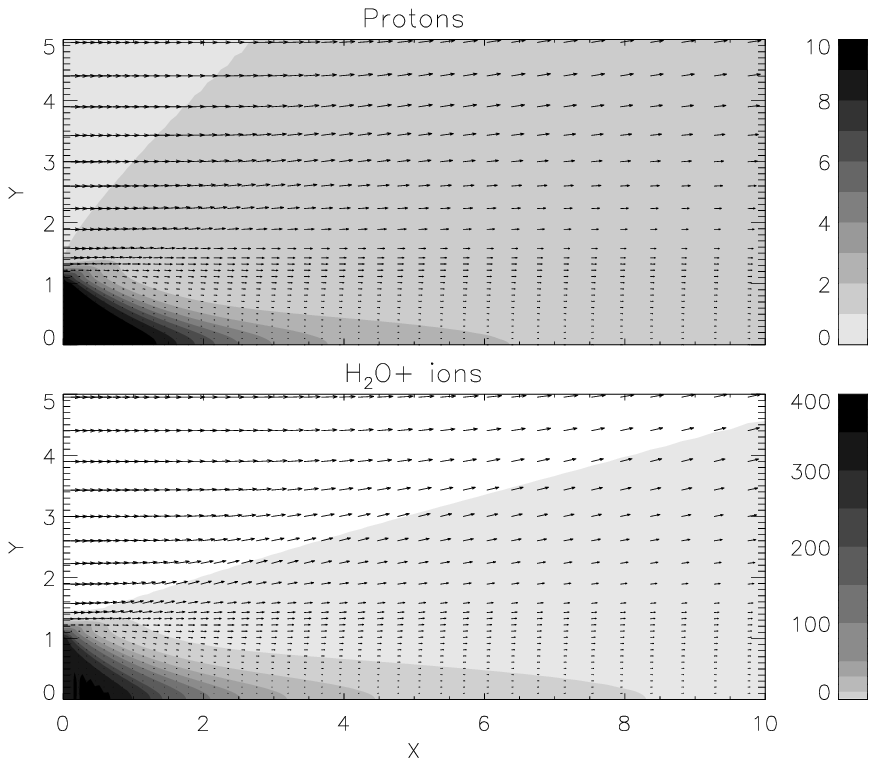}}
\caption {Same as in Figure 2 but for simulation Case 4
($R_{\rm eff}^{a,b} = 10$, $\nu_o = 0.1$).}
\label{fig6}
\end{figure}
%
%

\subsection{Effect of viscous-like forces}

\begin{figure}
\centering
\includegraphics{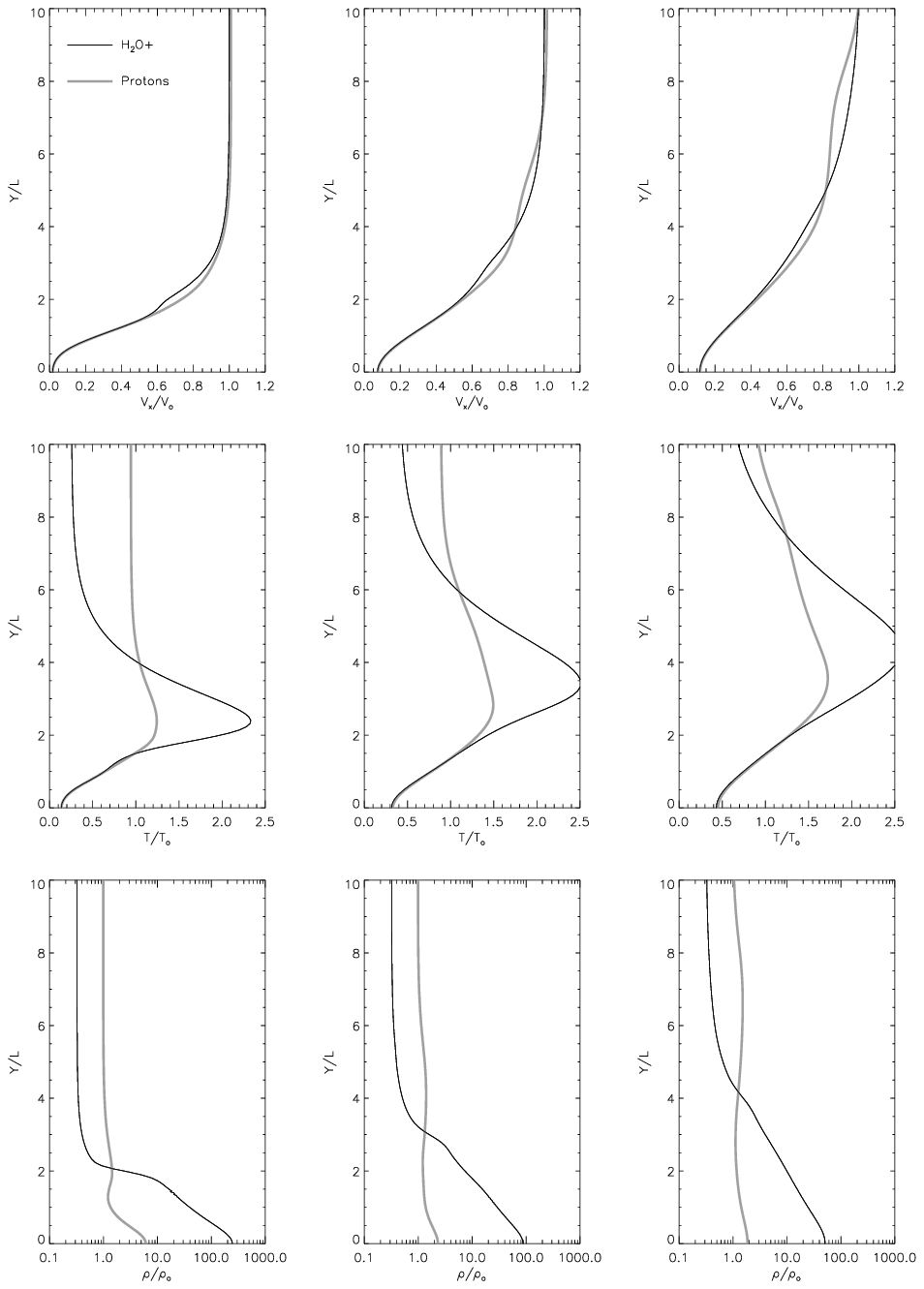}
\caption {Same as in Figure 3 but for simulation Case 4
($R_{\rm eff}^{a,b} = 10$, $\nu_o = 0.1$).}
\label{fig7}
\end{figure}

To analyse the effect of the viscous-like momentum transport between the solar 
wind and material in the comet's plasma tail, we compare 3 simulations with 
the same inter-species coupling parameter, $\nu_o = 0.1$, but different values 
of the effective Reynolds number. Figures \ref{fig6} and \ref{fig7} shown the resulting 
flow geometry and vertical profiles, respectively, for our Case 4, characterized 
by a higher viscosity corresponding to a lower effective Reynolds number, $R_{\rm eff}  = 10$, 
than Case 1. Comparing the global geometry of the flow in this case (Figure \ref{fig6}) 
with that in a case with greater Reynolds number, $R_{\rm eff}^{a,b} = 30$ 
(Figure \ref{fig2}) we see that after 1234 crossing 
times, the erosion of the tail is much greater in this high viscosity case 
for both species. This result is expected as well as the increase in the 
thickness of the boundary layer as we decrease the effective Reynolds number. This is
clearly seen when comparing the vertical profiles of the flow properties shown in
Figure \ref{fig3} (medium viscosity) and Figure \ref{fig7} (high viscosity). For example, 
as shown in Figure \ref{fig7} for $x = 5$, the top of the boundary layer 
increases from $y = 2.8$ for $Re^{a,b} = 30$ to approximately 
$y = 3.7$ for $R_{\rm eff}^{a,b} = 10$.
The increased thickness of the boundary layer as we decrease $R_{\rm eff}^{a,b}$, effectively 
represents a more blunt obstacle to the solar wind flow. Hence, the height of the 
boundary layer also increases as we decrease $Re^{a,b}$. This is also shown in Figure 
\ref{fig7} where, for example at $x = 5$, the height of the shock front is 
located approximately at $y = 7$; about 2 scale units higher than the shock
front location for
the model with lower effective viscosity (Figure \ref{fig3}). 

%
%
%
%

\begin{figure}
\resizebox{\hsize}{!}{\includegraphics{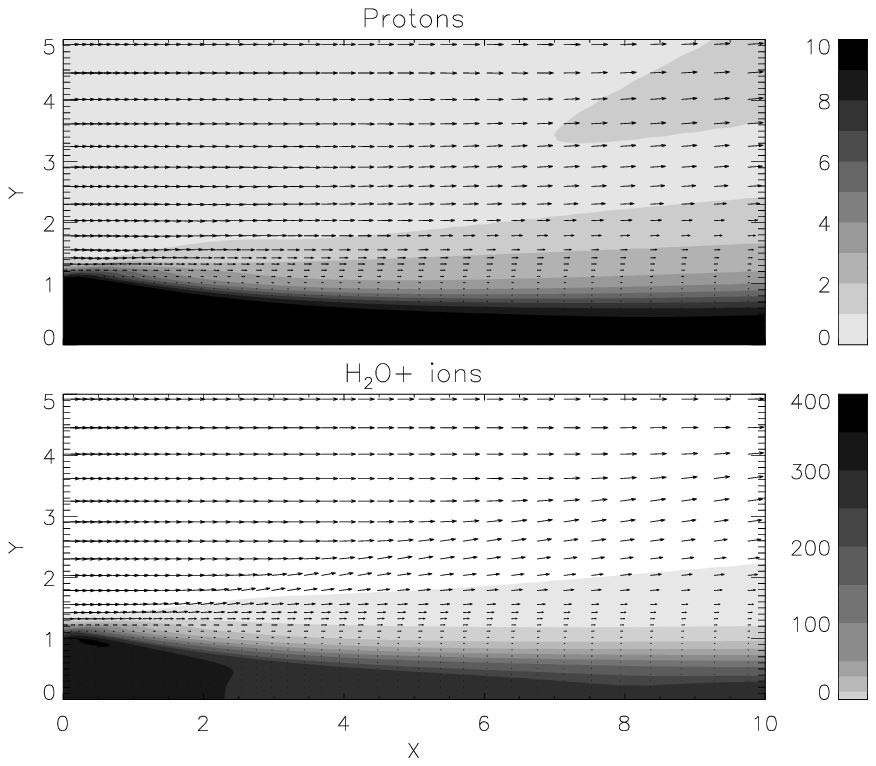}}
\caption {Same as in Figure 2 but for simulation Case 5
($R_{\rm eff}^{a,b} = 100$, $\nu_o = 0.1$).}
\label{fig8}
\end{figure}

\begin{figure}
\centering
\includegraphics{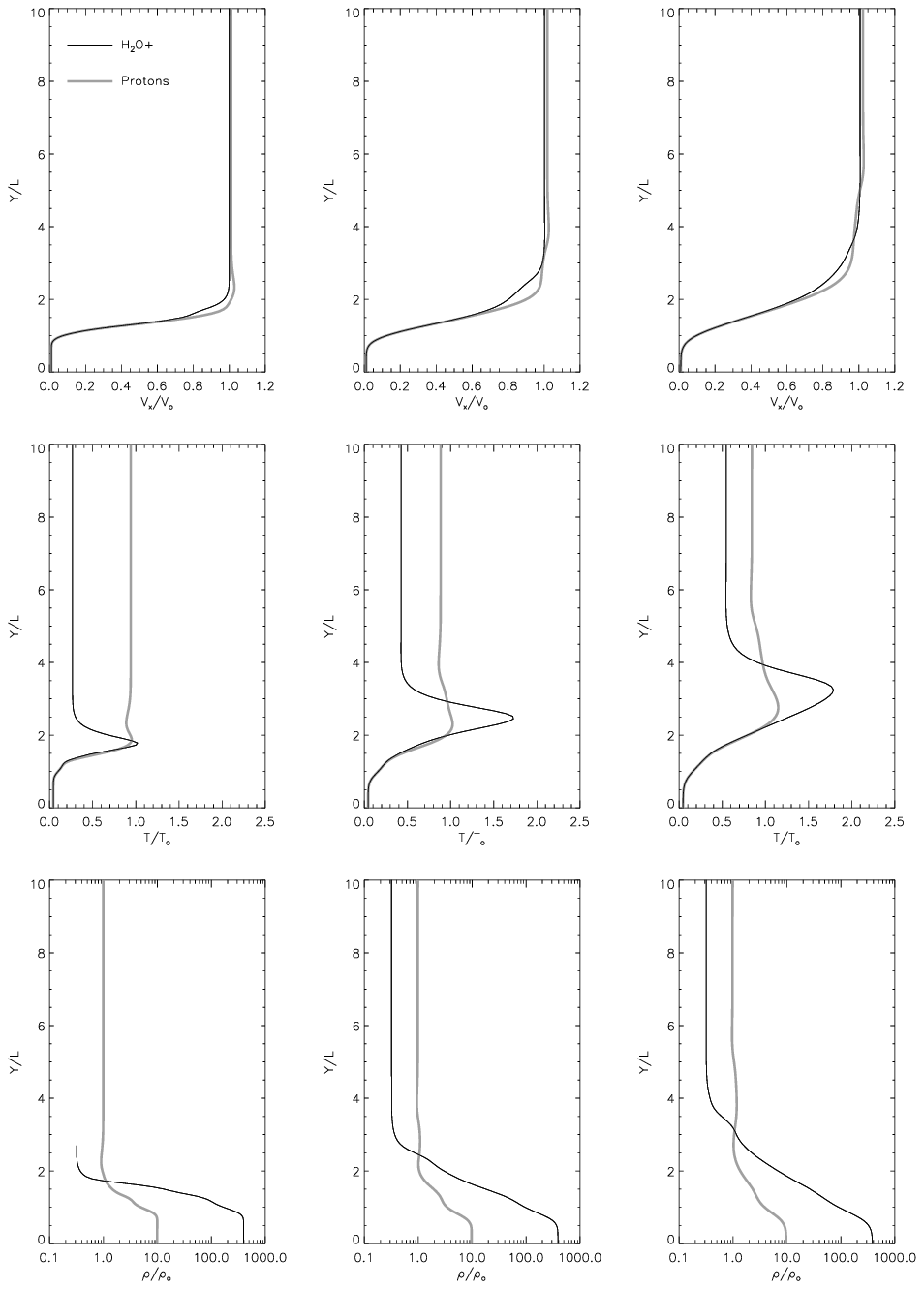}
\caption {Same as in Figure 3 but for simulation Case 5
($R_{\rm eff}^{a,b} = 100$, $\nu_o = 0.1$).}
\label{fig9}
\end{figure}

The tendency seen in going from high ($R_{\rm eff}^{a,b} = 10$) to medium 
effective viscosity ($R_{\rm eff}^{a,b} = 30$) is confirmed by 
comparing with results with an even smaller viscosity, such as Case 5 which 
corresponds to a model with $Re^{a,b} = 100$, shown in Figures \ref{fig8} and 
\ref{fig9}. As expected, a decreased viscosity leads to significantly less 
erosion of the tail than in Cases 1 and 4 (medium and high viscosity respectively) 
as shown in Figure \ref{fig8}. 
Also, as discussed above and as shown in Figure \ref{fig9}, the top of the 
boundary layer decreases as we increase the Reynolds number, and consequently 
the location of the shock front also decreases. 
For example, in the profiles corresponding 
to $x = 5$ in Figure \ref{fig9}, we find that the top of the boundary layer 
decreases from $y = 2.8$ for $R_{\rm eff}^{a,b} = 30$ to $y = 2.2$ for 
$R_{\rm eff}^{a,b} = 100$.
In regards to the location of the shock front, this 
goes from $y = 4.8$ for $R_{\rm eff}^{a,b} = 30$ to 
approximately $y = 3.7$ for $R_{\rm eff}^{a,b} = 100$.

%
%
%
%

\section{Discussion}
\label{sec:discussion}

In view of the uncertainty about the precise physical mechanisms
giving rise to the effective viscosity, we have assumed 
that the effective Reynolds 
number for both species is the same in the calculations presented above. 
However, we have also carried out simulations having distinct effective
Reynolds number for each species and find that for a medium value of 
the inter-species coupling, $\nu_o = 0.1$, the results are almost identical 
to those with a single value for the effective Reynolds number for both species 
(equal to the effective Reynolds number of species $b$). For example, for a 
case with $R_{\rm eff}^a = 100$, $R_{\rm eff}^b = 10$ and $\nu_o = 0.1$, 
the vertical profile 
of flow properties for the H$_2$O+ ions at all $x$ locations is almost identical 
to that shown in Figure \ref{fig7} for Case 4, characterized 
by $R_{\rm eff}^a = R_{\rm eff}^b = 10$
and $\nu_o = 0.1$. 
The vertical profile of flow properties for the protons, while not identical, is 
still very similar to Case 4.  This suggests that viscous stresses, particularly
in the species that dominates the mass of the problem, are the 
dominant factor in the flow dynamics.  

%
%
%
%

\begin{figure}
\centering
\includegraphics{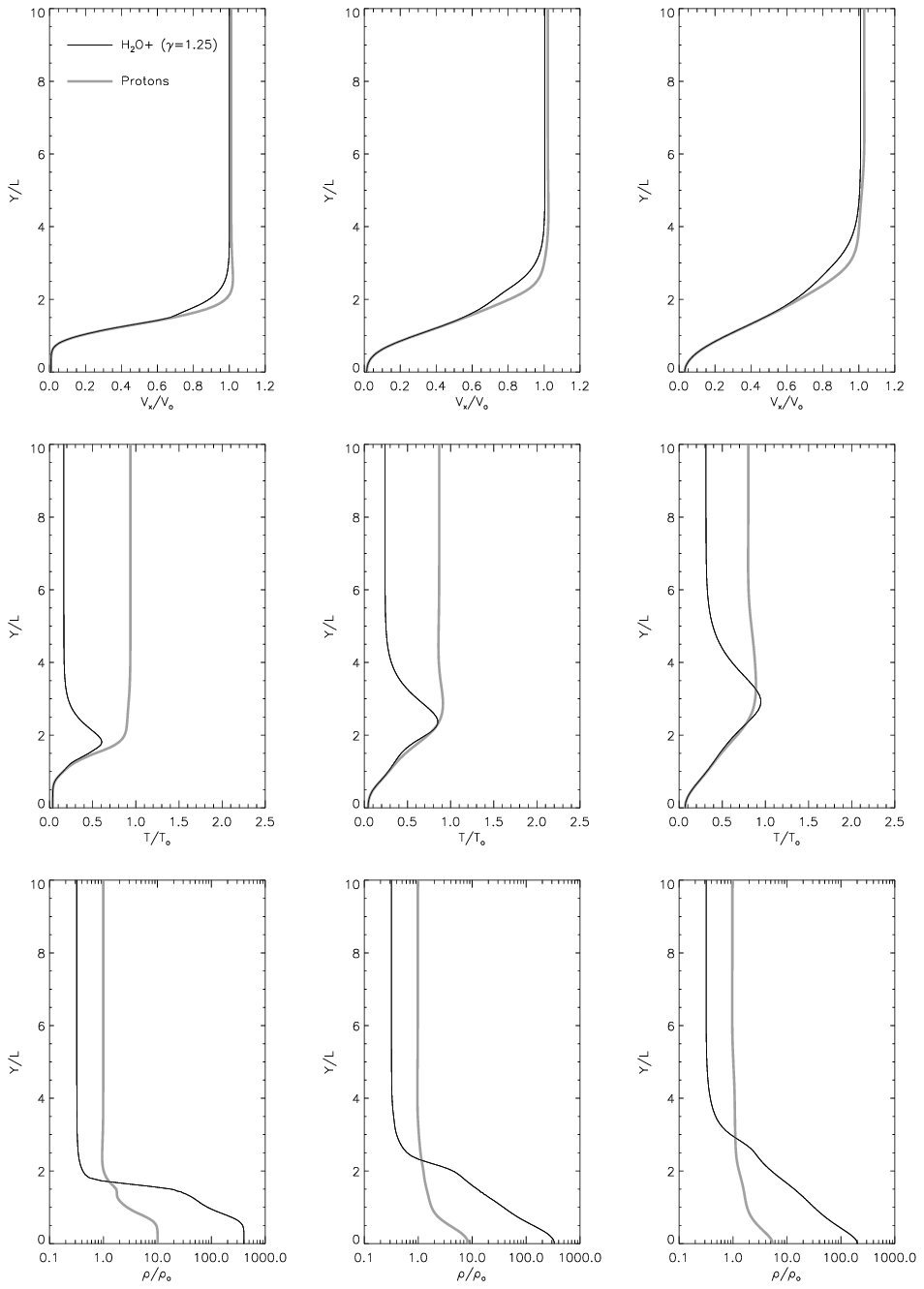}
\caption {Comparison of the vertical profiles of flow properties 
for a case characterized by the same value of $R_{\rm eff}$ and 
$\nu_o$ as Case 1, but with $\gamma^b = 1.25$. 
Profiles at $x = 2$ (left column of panels), 
$x=5$ (middle column) and $x=8$ (right column) are shown. 
Gray lines indicate the properties 
of the proton plasma and black lines denote the properties of 
the H$_2$O+ plasma. All quantities are in normalized units.}
\label{fig10}
\end{figure}

 As mentioned in section 2, in the results presented above we have assumed 
that the adiabatic index for both species is the same, $\gamma^a = \gamma^b = 1.67$.
While this value of $\gamma$ can be safely assumed for the solar wind plasma 
(assuming thermal equilibrium for the species), it is not so clearly valid for
the H$_2$O$^+$ plasma in which the excitation of rotational and vibrational 
degrees of freedom may lead to a lower value of $\gamma$ (again assuming 
thermal equilibrium for the species). In order to illustrate the effect of 
a different, lower value of the adiabatic index for cometary plasma, we  
have also conducted simulations with a value $\gamma^b = 1.25$ for the 
adiabatic index of the H$_2$O$^+$ plasma. This value corresponds to a 
gas composed of triatomic molecules in thermal equilibrium at a high enough
temperature for all molecular degrees of freedom to be excited. 
Results for this case, $\gamma^a = 1.67$ and $\gamma^b = 1.25$, with 
the same effective viscosity and interspecies coupling parameters as Case 1 
($R_{\rm eff}^{ab} = 30$ and $\nu_o = 0.1$) are shown in Figure \ref{fig10}
which shows the vertical profiles of $V_x$, $T$ and $\rho$ for both 
species in both cases.

%
%
%
%

 Clearly evident when comparing Figure \ref{fig10} ($\gamma^b = 1.25$) 
and Figure \ref{fig3} ($\gamma^b = 1.67$) is the fact that if the cometary 
plasma is characterized by a lower value of the adiabatic index, 
the heating of the H$_2$O$^+$ plasma is significantly 
reduced in the boundary layer, since part of the dissipated energy goes 
to the excitation of the additional degrees of freedom corresponding 
to the lower value of $\gamma$. This leads to less plasma expansion in 
the region and a thinner velocity boundary layer. The height of the shock front 
is consequently reduced. In future contributions we shall address the 
issue of the appropriate value of $\gamma$ for the cometary plasma.

\subsection{Comparison with {\em in situ} measurements}

A comparison of our results with the {\em in situ} measurements 
made by the Giotto spacecraft, as it flew by comet Halley in March of 1986, is 
not straightforward. The simplified geometry we are using in our 
simulations to study the interaction 
in the tail region exclusively, precludes a direct
comparison. Nevertheless, some insight into the implications 
of our results can be obtained from a simplified comparison.  

 Once a particular application scenario has been chosen, values for the 
characteristic length, velocity, density and temperature used in 
the adimensionalization of the equations of motion (section \ref{sec:problem}) 
can be established. For comet Halley, using the {\em in situ} measurements 
reported in Goldstein et al. (1986), Johnstone et al. (1986) and Perez-de-Tejada (1989),
we adopt $L$ = 150,000 km, $V_o$ = 250 km/s, $\rho_o$ = 1.67 $\times 10^{-23}$  
gm/cm$^3$ and $T_o$ = 2.5 $\times 10^5$ K.

\begin{figure}
\resizebox{\hsize}{!}{\includegraphics{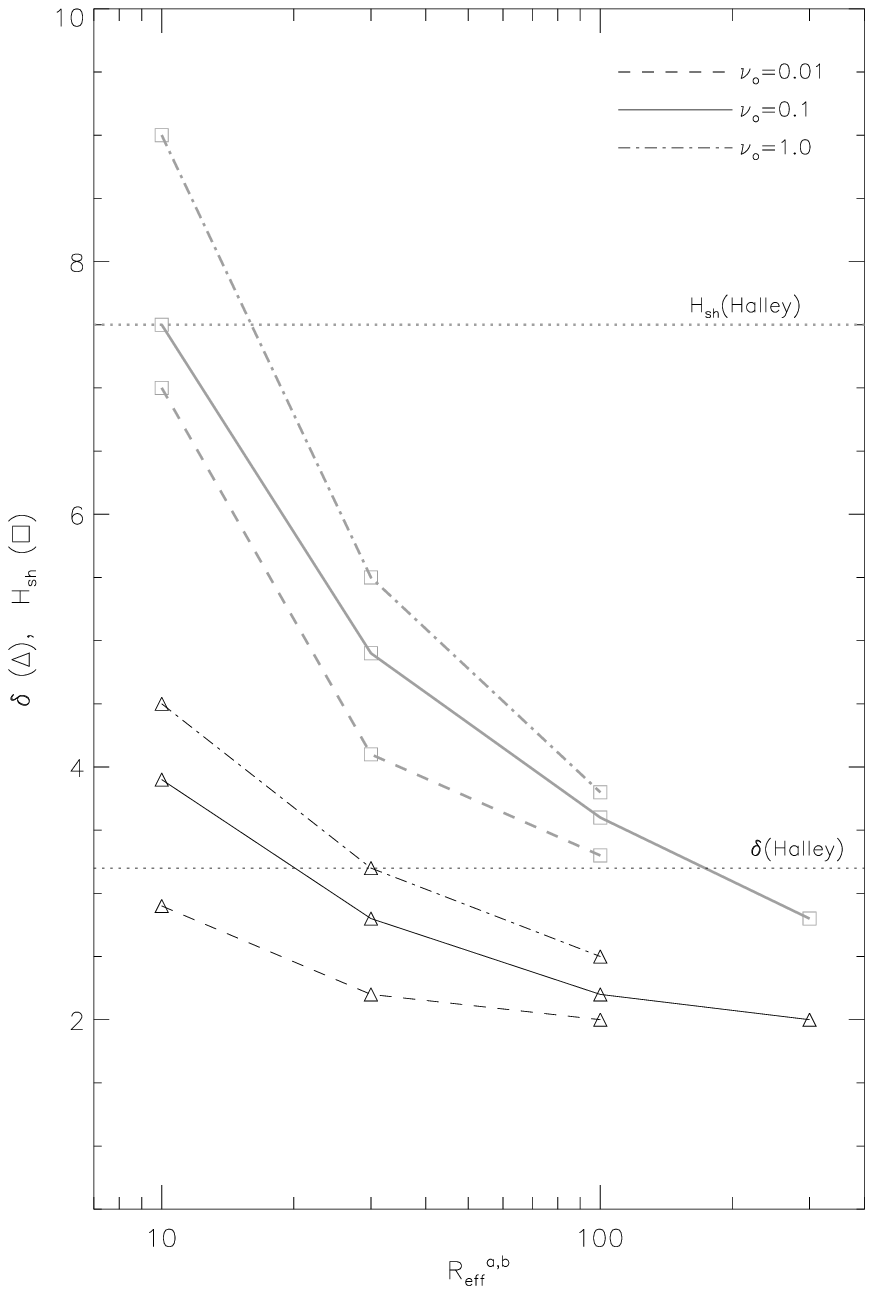}}
\caption {Height of the shock front, $H_{sh}$ (gray lines with 
squares) and thickness of the boundary layer, $\delta$ 
(black lines with triangles), as a function of effective 
Reynolds number, for a set of models with
different value for the inter-species coupling parameter, $\nu_o$. Values 
for these scale-heights correspond to $x = 5$ in our model. The dotted 
lines indicate the height of the shock front (gray) and the location of the 
intermediate transition (black) during the inbound portion of Giotto's flyby 
through the tail of comet Halley.}
\label{fig11}
\end{figure}

According to Johnstone et al. (1986), the Giotto spacecraft observed 3 distinct 
transitions in the plasma properties on its inbound trajectory towards comet
Halley's nuclear region: (1) The outermost transition occurs about 900000 km from 
the point of closest approach and can be identified as the 
bow shock crossing. (2) The cometopause, where the density of cometary ions sharply 
increases, can be located at around 150000 km from closest approach
(Perez-de-Tejada, 1989). (3) Approximately midway between the shock location and 
the cometopause, at about 400000 km from closest approach, the so called intermediate
transition signals the top of the viscous boundary layer according to 
the viscous flow interpretation of the solar-wind-comet interaction 
given by Perez-de-Tejada (1989). Pending a more 
detailed comparison of the Giotto measurements with the results of our simulations,
which should take into account the full geometry of the problem, let us
identify the cometopause detected in the measurements with the region of 
very strong H$_2$O+ density increase in our simulations, located  
at $y = 1.0$ (approximately) in our normalized units. 
Under this assumption, in Figure \ref{fig11}
we compare the thickness of the boundary layer and the height of the shock 
front evaluated from our simulation results at $x = 5$, for models with 
different effective Reynolds number ($R_{\rm eff}^{a,b}$) 
and inter-species coupling parameter ($\nu_o$). As 
already seen, both the thickness of the boundary layer and the height of the 
shock front decrease with increasing Reynolds number so that, almost irrespective
of the value of $\nu_o$, a low value of $R_{\rm eff}^{a,b}$ is required to explain the 
measured transition locations. 

%
%
%
%

Also evident in Figure \ref{fig11} is the 
dependence of the transition locations on the value of the inter-species 
coupling parameter. In simulations
with a strong inter-species coupling, the solar wind ions 
are able to transfer momentum to cometary ions more 
efficiently giving rise to a thicker boundary layer and
higher shock front. The opposite is true when 
both species are weakly coupled ($\nu_o = 0.01$). 
In such case solar wind ions flow by cometary 
plasma interacting very weakly. Less momentum is 
transferred between the solar wind and cometary plasma 
in a situation reminiscent of a high Reynolds number case. 
Our analysis of scale-heights
is based not only on the properties of the velocity profiles in our simulations. 
As pointed out by Perez-de-Tejada (1989), there are corresponding 
changes in the density and temperature of the gas as one enters a boundary 
layer. The heating and expansion characteristic of 
viscous boundary layers are also found in our results, particularly
for cases with low Reynolds number and high inter-species 
coupling parameter..

It is worth mentioning that somewhat similar properties of the flow were also
measured by the ICE spacecraft in its flyby through comet Giacobinni-Zinner
as discussed in Ip (2004). A comparison of the location of the transition from 
the sheath region to the so-called transition region and the bow shock location
as estimated by Reme (1991), corrected for the different height of the 
cometopause, yields very similar relative positions to those shown by the 
dotted lines in Figure \ref{fig11} for the transitions in comet Halley. 
In future work we will address the differences in the flow properties measured 
in comet Halley and in comet Giacobinni-Zinner.  

\subsection{Implications for 3D geometry}

It is important to emphasize that the geometry presented in this paper is 
derived from a 2D model. In Venus, as discussed by Perez-de-Tejada 
(1995), the viscous-like interaction between solar wind and ionospheric plasmas  
takes place preferentially over the magnetic poles of the planet (defined in terms
of the incident IMF), where the pile-up of magnetic field lines is 
less than around equatorial latitudes. According to Perez-de-Tejada (1995),
up to about 80$^{\rm o}$ SZA, the piled-up magnetic field over the 
dayside ionosphere and along the flanks, inhibits in some degree a direct, 
viscous-like interaction 
between solar wind and ionospheric plasmas. 

If we apply these ideas to the solar wind-comet interaction, this implies 
that the flow properties we have computed here, correspond more closely
to locations 
over and downstream from the magnetic poles of the comet. For different 
locations along the tail, the piled-up magnetic field may prevent an 
efficient viscous-like dragging of ionospheric material, ${\bf J} \times {\bf B}$ 
forces may be more important and the flow dynamics may be 
better modeled in terms of an MHD model as those of Wegmann (2002) and
Jia et al. (2007). As the IMF is constantly changing direction on a wide 
range of amplitudes and timescales, the region of viscous-like interaction 
between the solar wind and cometary plasma, changes with time. Given
the typical IMF orientation is approximately in the ecliptic plane, one 
should expect that the flow within +/- 20$^{\rm o}$, measured in the 
$y$-direction (as typically defined) from the magnetic poles of the comet, 
is best described by our model.     

\section{Conclusions}

We have presented results for the numerical simulation of the interaction between
the solar wind and the plasma in the tail of a comet, taking into account the effect
of viscous-like stresses previously argued to be important by Perez-de-Tejada et al
(1980). To our knowledge, this is the first time that viscous-like effects have been 
incorporated into such studies. 
Our results indicate the existence of 3 distinct transitions in the flow properties:
outermost we find a shock front, innermost we have the cometopause and an 
intermediate transition which we can identify with the height of the boundary layer 
characterized by a fast decline in the anti-sunward flow velocity, and the onset of 
plasma heating and expansion due to viscous-like dissipation. The location of these 
transitions depends on the flow parameters, namely the effective Reynolds number of
the flow for each species, $R_{\rm eff}^{a,b}$, 
and the inter-species coupling parameter, $\nu_o$. 

By comparing the flow properties from our numerical 
simulations to the location of the shock front and intermediate transition,
as measured by the Giotto spacecraft as it approached the nucleus of
comet Halley, we find that, almost irrespective of the 
strength of the inter-species coupling, $\nu_o$; a low value of the 
effective Reynolds number, 
approximately $R_{\rm eff}^{a,b} \lesssim 20$ for both species, is required to reproduce the 
measured transition locations. This implies, in the context of our model, 
that the measured flow 
properties cannot be explained if one does not take into account the 
viscous-like forces in the interaction of the solar wind and the plasma tail of a 
comet. Although the conclusions drawn from this study are
strictly applicable only to comet Halley and solar wind conditions
at the time the {\em in situ} measurements were taken, one may 
speculate that viscous-like processes may be important in the
solar wind-comet interaction in general. 

 It is important to emphasize that, this being the first 
attempt to include viscous-like
forces in the numerical simulation of the interaction of the solar wind
with a comet's plasma environment, there are many pending issues
still to be addressed that could have potentially important 
consequences on the details of the solutions obtained under our 
simplified treatment. First and foremost, the precise forms we
are using for the viscous like stress and effective interspecies coupling, 
may be questioned. As we have argued in the Introduction, 
plasma properties imply that ``normal'' viscosity is 
negligible in the region under consideration. Hence, we are invoking an 
effective viscosity presumably resulting from plasma turbulence and/or
wave-particle interactions. However, the precise form of the
terms corresponding to viscous-like momentum transfer in the 
equations of motion (Bousinessq hypothesis) is not formally 
demonstrated. Also, as we have discussed in the Formulation section
of the paper, the interspecies coupling terms we are
using can not be strictly derived for a plasma as the one we are modelling.  
In view of these arguments, one may consider that the work reported in this
paper is only an academic exercise of 
questionable applicability to the problem of solar wind-cometary plasma 
interaction. In such case, a similar conclusion must be reached in 
regards to many other studies of fluid dynamics that use 
similar approaches to modelling effective viscosity and interspecies
coupling. 

Additional important effects still to be considered are the 
following: geometrical 
effects due to the curvature of the ionosphere are required 
for a more direct, quantitative comparison between {\em in situ}
measurements by the Giotto spacecraft and the results of simulations;
the interaction of the charged species with neutral gas ejected
from the comet which, especially in the vicinity of the nucleus,
is the most abundant species;
the effect of the magnetic field on the 
flow (particularly in the dayside and around the midplane of the near-tail 
region), 3D effects, incoming flow time dependence, etc. We  
believe that the further assessment of the relevance of these factors 
is beyond the present study. They are the subject of work
currently in progress and will be reported in future contributions.

\begin{acknowledgements}
      MRR acknowledges the support of research grant IN109409 of DGAPA-UNAM. 
HA acknowledges the support of research grant IN121406 of DGAPA-UNAM 
and CONACYT-M\'exico grant No. 508807.

\end{acknowledgements}

\end{document}